\documentclass[titlepage,10pt]{article}

\usepackage{amstext, amssymb, amsthm,natbib, amsmath,graphicx,rotating,array,booktabs}
\usepackage{algorithmic,algorithm}
\usepackage{geometry}

\newtheorem{theorem}{Theorem}[section]

\newtheorem{proposition}[theorem]{Proposition}

\newcommand{\svskip}{\vspace{1.75mm}}

\newcommand{\bb}{\boldsymbol{b}}
\newcommand{\bc}{\boldsymbol{c}}
\newcommand{\bd}{\boldsymbol{d}}
\newcommand{\be}{\boldsymbol{e}}

\newcommand{\br}{\boldsymbol{r}}
\newcommand{\bs}{\boldsymbol{s}}
\newcommand{\bt}{\boldsymbol{t}}
\newcommand{\bu}{\boldsymbol{u}}
\newcommand{\bv}{\boldsymbol{v}}
\newcommand{\bw}{\boldsymbol{w}}
\newcommand{\bx}{\boldsymbol{x}}
\newcommand{\by}{\boldsymbol{y}}
\newcommand{\bz}{\boldsymbol{z}}
\newcommand{\bA}{\boldsymbol{A}}
\newcommand{\bB}{\boldsymbol{B}}
\newcommand{\bC}{\boldsymbol{C}}
\newcommand{\bD}{\boldsymbol{D}}
\newcommand{\bE}{\boldsymbol{E}}
\newcommand{\bG}{\boldsymbol{G}}
\newcommand{\bI}{\boldsymbol{I}}
\newcommand{\bP}{\boldsymbol{P}}
\newcommand{\bQ}{\boldsymbol{Q}}

\newcommand{\bR}{\boldsymbol{R}}
\newcommand{\bX}{\boldsymbol{X}}
\newcommand{\bU}{\boldsymbol{U}}
\newcommand{\bV}{\boldsymbol{V}}
\newcommand{\bW}{\boldsymbol{W}}
\newcommand{\bY}{\boldsymbol{Y}}

\newcommand{\bbeta}{\boldsymbol{\beta}}

\newcommand{\btheta}{\boldsymbol{\theta}}

\newcommand{\blambda}{\boldsymbol{\lambda}}
\newcommand{\bPsi}{\boldsymbol{\Psi}}
\newcommand{\bxi}{\boldsymbol{\xi}}

\title{A Path Algorithm for Constrained Estimation}
\author{Hua Zhou \\
Department of Statistics \\
North Carolina State University \\
Raleigh, NC 27695-8203 \\
Phone: 919-515-2570 \\
E-mail: hua\_zhou@ncsu.edu \\
\\
Kenneth Lange \\
Departments of Biomathematics, \\
Human Genetics, and Statistics \\
University of California    \\
Los Angeles, CA 90095-1766 \\
Phone: 310-206-8076 \\
E-mail: klange@ucla.edu \\
}

\begin{document}
\maketitle

\baselineskip=20pt

\begin{abstract}
Many least squares problems involve affine equality and inequality constraints.  Although there are variety of methods for solving such problems, most statisticians find constrained estimation challenging. The current paper proposes a new path following algorithm for quadratic programming based on exact penalization. Similar penalties arise in $l_1$ regularization in model selection.  Classical penalty methods solve a sequence of unconstrained problems that put greater and greater stress on meeting the constraints.  In the limit as the penalty constant tends to $\infty$, one recovers the constrained solution.  In the exact penalty method, squared penalties are replaced by absolute value penalties, and the solution is recovered for a finite value of the penalty constant.  The exact path following method starts at the unconstrained solution and follows the solution path as the penalty constant increases. In the process, the solution path hits, slides along, and exits from the various constraints.  Path following in lasso penalized regression, in contrast, starts with a large value of the penalty constant and works its way downward. In both settings, inspection of the entire solution path is revealing. Just as with the lasso and generalized lasso, it is possible to plot the effective degrees of freedom along the solution path.  For a strictly convex quadratic program, the exact penalty algorithm can be framed entirely in terms of the sweep operator of regression analysis.  A few well chosen examples illustrate the mechanics and potential of path following.
\end{abstract}
\vspace{.1in}

\section{Introduction}

When constraints appear in estimation by maximum likelihood or least squares estimation, statisticians typically resort to sophisticated commercial software or craft specific optimization algorithms for specific problems. In this article, we develop a simple path algorithm for a general class of constrained estimation problems, namely quadratic programs with affine equality and inequality constraints. Besides providing constrained estimates, our new algorithm also delivers the whole solution path between the unconstrained and the constrained estimates. This is particularly helpful when the goal is to locate a solution between these two extremes based on criteria such as prediction error in cross-validation.

In recent years several path algorithms have been devised for specific $l_1$ regularization problems. The solution paths generated  vividly illustrate the tradeoffs between goodness of fit and sparsity. For example, a modification of the least angle regression (LARS) procedure can handle lasso penalized regression \citep{EfronHastieIainTibshirani04LARS}. \cite{RossetZhu07Path} give sufficient conditions for a solution path to be piecewise linear and expand its applications to a wider range of loss and penalty functions. \cite{Friedman08GPS} derives a path algorithm for any objective function defined by the sum of a convex loss and a separable penalty (not necessarily convex).  The separability restriction on the penalty term excludes many of the problems studied here.  \cite{TibshiraniTaylor10GenLasso} devise a path algorithm for generalized lasso problems. Their formulation is similar to ours, but there are two fundamental differences. First, inequality constraints are excluded in their formulation. Our new path algorithm handles both equality and inequality constraints gracefully. Second, they pass to the dual problem and then translate the solution path of the dual problem back to the solution path of the primal problem. In our view, attacking the primal problem directly leads to a simpler algorithm, indeed one driven entirely by the classical sweep operator of regression analysis. These gains in conceptual clarity and implementation ease constitute major pluses for statisticians.  As we will show, the degrees of freedom formula derived for the lasso \citep{EfronHastieIainTibshirani04LARS,ZouHastieTibshirani07LassoDF} and generalized lasso \citep{TibshiraniTaylor10GenLasso} apply equally well in the presence of inequality constraints.

Our object of study will be minimization of the quadratic function
\begin{eqnarray}
f(\bx) & = &  \frac{1}{2} \bx^t \bA \bx + \bb^t \bx + c \label{quadratic_function}
\end{eqnarray}
subject to the affine equality constraints $\bV\bx = \bd$ and the affine inequality constraints $\bW \bx \le \be$. Throughout our discussion we assume that the feasible region is nontrivial and that the minimum is attained. If the symmetric matrix $\bA$ has a negative eigenvalue $\lambda$ and corresponding unit eigenvector $\bu$, then $\lim_{r \to \infty} f(r\bu) = -\infty$ because the quadratic term $\frac{1}{2} (r \bu)^t \bA (r \bu) = \frac{\lambda}{2}r^2$ dominates the linear term $r \bb^t \bu$. To avoid such behavior, we initially assume that all eigenvalues of $\bA$ are positive. This makes $f(\bx)$ strictly convex and coercive and guarantees a unique minimum point subject to the constraints. In linear regression $\bA = \bX^t \bX$ for some design matrix $\bX$. In this setting $\bA$ is positive definite provided $\bX$ has full column rank.  The latter condition is only possible when the number of cases equals or exceeds the number of predictors.  If $\bA$ is positive semidefinite and singular, then adding a small amount of ridge regularization $\epsilon \bI$ to it can be helpful \citep{TibshiraniTaylor10GenLasso}. Later we indicate how path following extends to positive semidefinite or even indefinite matrices $\bA$.

In multi-task models in machine learning, the response is a $d$-dimensional vector $\bY \in \mathbb{R}^d$, and one minimizes the squared Frobenius deviation
\begin{eqnarray}
    \frac 12 \|\bY - \bX \bB\|_{\text{F}}^2 \label{eqn:multi-task}
\end{eqnarray}
with respect to the $p \times d$ regression coefficient matrix $\bB$. When the constraints take the form $\bV \bB \le \bD$ and
$\bW \bB = \bE$, the problem reduces to quadratic programming as just posed. Indeed, if we stack the columns of $Y$ with the $\text{vec}$ operator, then the problem reduces to minimizing $\frac{1}{2} \|\text{vec}(\bY) - \bI \otimes \bX \text{vec}(\bB)\|_2^2$.  Here  the identity
$ \text{vec}(\bX \bB) = \bI \otimes \bX \text{vec}(\bB)$ comes into play involving the Kronecker product and the identity matrix $\bI$.   The same identity allows to rewrite the constraints as $ \bI \otimes \bV \text{vec}(\bX) = \text{vec}(\bD)$ and $\bI \otimes \bW \text{vec}(\bX) \le \text{vec}(\bE)$.

\begin{figure}
$$
\begin{array}{cc}
\includegraphics[width=2.5in]{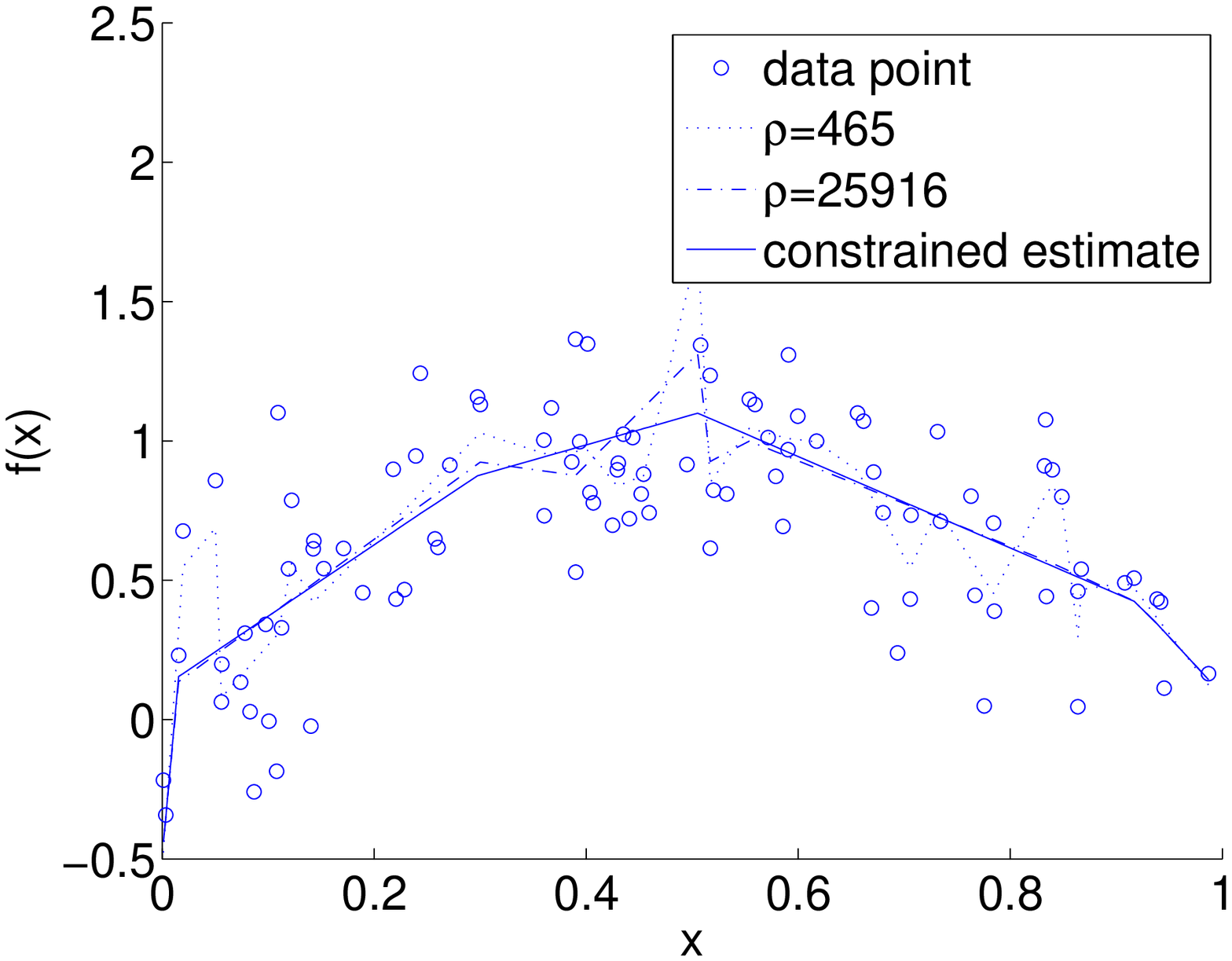} & \includegraphics[width=2.5in]{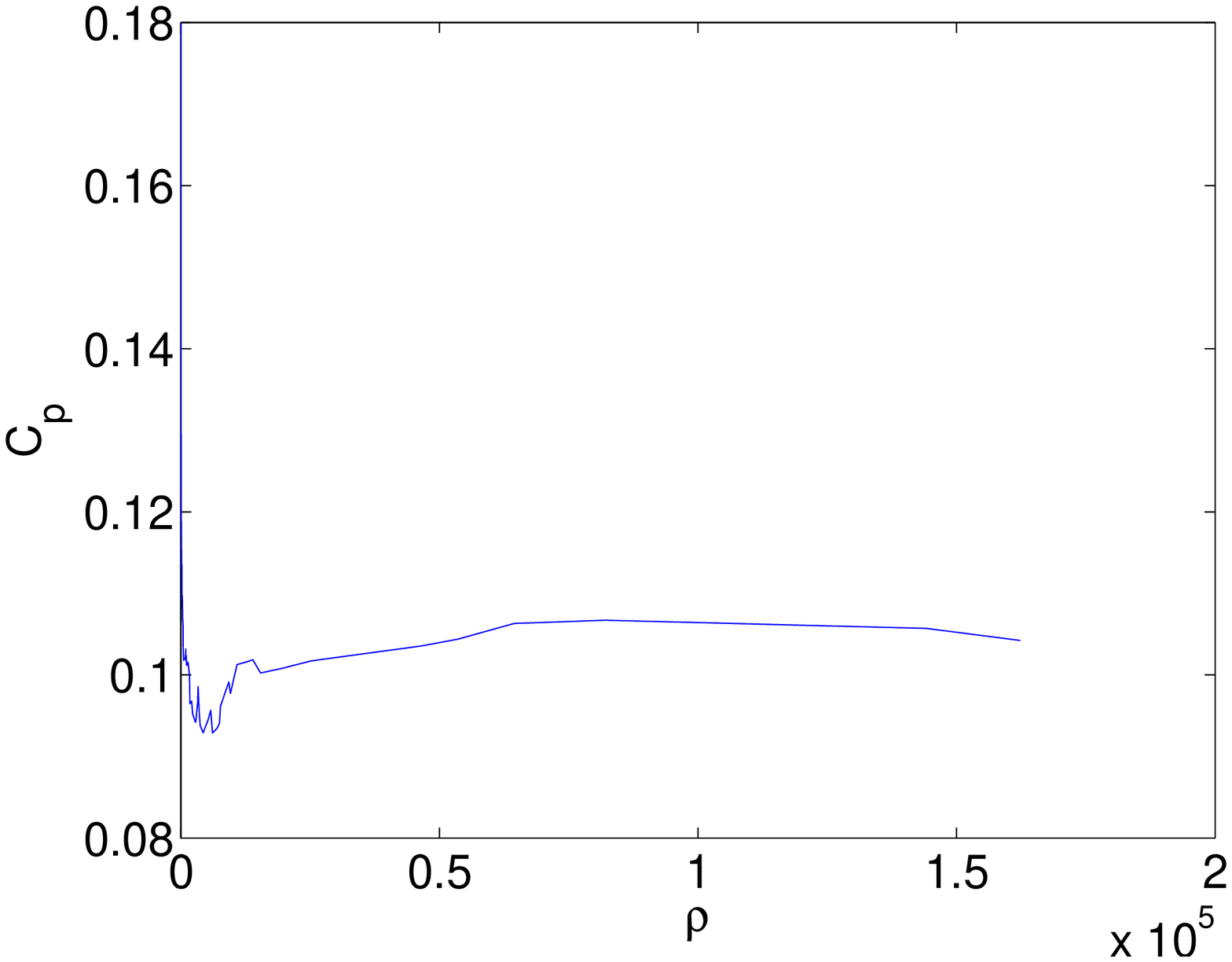}
\end{array}
$$
\caption{Path solutions to the concave regression problem. Left: the unconstrained solution (original data points), two intermediate solutions (dotted and dashed lines), and the concavity constrained solution (solid line). Right: the $C_p$ statistic as a function of the penalty constant $\rho$ along the solution path.}
\label{fig:concreg}
\end{figure}

As an illustration, consider the classical concave regression problem \citep{Hildreth54ConcaveReg}. The data consist of a scatter plot $(x_i,y_i)$ of $n$ points with associated weights $w_i$ and predictors $x_i$ arranged in increasing order. The concave regression problem seeks the estimates $\theta_i$ that minimize the weighted sum of squares
\begin{eqnarray}
    \sum_{i=1}^n w_i (y_i - \theta_i)^2 \label{eqn:concave-reg}
\end{eqnarray}
subject to the concavity constraints
\begin{eqnarray}
    \frac{\theta_i - \theta_{i-1}}{x_i - x_{i-1}}  &\ge & \frac{\theta_{i+1} - \theta_i}{x_{i+1}-x_i}, \hspace{.15in} i=2,\ldots,n-1.
    \label{eqn:concave-constraints}
\end{eqnarray}
The consistency of concave regression is proved by \cite{HansonPledger76ConcaveReg}; the asymptotic distribution of the estimates and their rate of convergence are studied in subsequent papers \citep{Mammen91Nonparam,GroeneboomJongbloedWellner01ConvexReg}. Figure \ref{fig:concreg} shows a scatter plot of 100 data points. Here the $x_i$ are uniformly sampled from the interval [0,1], the weights are constant, and $y_i = 4x_i(1-x_i) + \epsilon_i$, where the $\epsilon_i$ are i.i.d.\ normal with mean 0 and standard deviation $\sigma=0.3$. The left panel of Figure \ref{fig:concreg} gives four snapshots of the solution path. The original data points $\hat \theta_i=y_i$ provide the unconstrained estimates. The solid line shows the concavity constrained solution. The dotted and dashed lines represent intermediate solutions between the unconstrained and constrained solutions. The degrees of freedom formula derived in Section \ref{sec:dof} is a vehicle for model selection based on criterion such as $C_p$, AIC, and BIC. For example, the $C_p$ statistic
\begin{eqnarray*}
    C_p(\hat{\btheta}) & = & \frac 1n \|\by - \hat{\btheta}\|_2^2 +  \frac 2n \sigma^2 \text{df}(\hat{\btheta})
\end{eqnarray*}
is an unbiased estimator of the true prediction error \citep{Efron04CV} under the estimator $\hat{\btheta}$. The right panel shows the
$C_p$ statistic along the solution path. In this example the design matrix is a diagonal matrix. As we will see in Section \ref{sec:examples}, postulating a more general design matrix or other kinds of constraints broadens the scope of applications of the path algorithm and the estimated degrees of freedom.

Here is a roadmap to the remainder the current paper. Section \ref{sec:exact-penalty-method} reviews the exact penalty method for optimization and clarifies the connections between constrained optimization and regularization in statistics. Section \ref{sec:path-algorithm} derives in detail our path algorithm. Its implementation via the sweep operator and QR decomposition are described in Sections \ref{sec:implementation} and \ref{path_extensions_section}. Section \ref{sec:dof} derives the degrees of freedom formula. Section \ref{sec:examples} presents various numerical examples. Finally, Section \ref{sec:conclusions} discusses the limitations of the path algorithm and hints at future generalizations.

\section{The Exact Penalty Method}
\label{sec:exact-penalty-method}

Exact penalty methods minimize the function
\begin{eqnarray*}
{\cal E}_{\rho}(\bx) & = & f(\bx)+\rho \sum_{i=1}^r |g_i(\bx)|+\rho \sum_{j=1}^s \max\{0,h_j(\bx)\},
\end{eqnarray*}
where $f(\bx)$ is the objective function, $g_i(\bx)=0$ is one of $r$ equality constraints,
and $h_j(\bx) \le 0$ is one of $s$ inequality constraints.  It is interesting to compare
this function to the Lagrangian function
\begin{eqnarray*}
{\cal L}(\bx) & = & f(\bx)+ \sum_{i=1}^r \lambda_i g_i(\bx)+\sum_{j=1}^s \mu_j h_j(\bx)
\end{eqnarray*}
that captures the behavior of $f(\bx)$ at a constrained local minimum $\by$.
By definition the Lagrange multipliers satisfy the conditions $\nabla {\cal L}(\by) ={\bf 0}$
and $\mu_j \ge 0$ and $\mu_j h_j(\by)=0$ for all $j$.  In the exact
penalty method we take
\begin{eqnarray}
\rho & > & \max \{|\lambda_1|,\ldots,|\lambda_r|,\mu_1,\ldots,\mu_s\}. \label{big_rho}
\end{eqnarray}
This choice creates the majorization $f(\bx) \le {\cal E}_{\rho}(\bx)$
with $f(\bz)={\cal E}_{\rho}(\bz)$ at any feasible point $\bz$.  Thus,
minimizing ${\cal E}_{\rho}(\bx)$ forces $f(\bx)$ downhill. Much
more than this is going on however.  As the next proposition proves,
minimizing ${\cal E}_{\rho}(\bx)$ effectively minimizes $f(\bx)$
subject to the constraints.

\begin{proposition} \label{proposition1}
Suppose the objective function $f(\bx)$ and the constraint functions
are twice differentiable and satisfy the Lagrange multiplier rule at the
local minimum $\by$. If inequality (\ref{big_rho}) holds and
$\bv^*d^2{\cal L}(\by)\bv>0$ for every vector $\bv \ne {\bf 0}$ satisfying
$dg_i(\by)\bv=0$ and $dh_j(\by)\bv \le 0$ for all active inequality constraints, then
$\by$ furnishes an unconstrained local minimum of ${\cal E}_{\rho}(\bx)$.
If $f(\bx)$ is convex, the $g_i(\bx)$ are affine, the $h_j(\bx)$ are
convex, and Slater's constraint qualification holds, then $\by$ is
a minimum of ${\cal E}_{\rho}(\bx)$ if and only if $\by$ is a minimum
of $f(\bx)$ subject to the constraints.  In this convex programming context, no
differentiability assumptions are needed.
\end{proposition}
{\bf Proof:} The conditions imposed on the quadratic form
$\bv^*d^2{\cal L}(\by)\bv>0$ are well-known sufficient conditions
for a local minimum.  Theorems 6.9 and 7.21 of the reference \citep{Ruszczynski06Book} prove all of the foregoing assertions. \qed \svskip

\section{The Path Following Algorithm}
\label{sec:path-algorithm}

We now resume our study of minimizing the objective function (\ref{quadratic_function}) subject to the affine equality constraints $\bV\bx = \bd$ and the affine inequality constraints $\bW \bx \le \be$.  The corresponding penalized objective function takes the form
\begin{eqnarray}
{\cal E}_\rho(\bx) & = & \frac{1}{2} \bx^t \bA \bx + \bb^t \bx + c + \rho \sum_{i=1}^r |\bv_i^t \bx - d_i| + \rho \sum_{j=1}^s (\bw_j^t \bx - e_i)_+.   \label{eqn:path-penalized-obj}
\end{eqnarray}
Our assumptions on $\bA$ render ${\cal E}_\rho(\bx)$ strictly convex and coercive and guarantee a unique minimum point $\bx(\rho)$. The generalized lasso problem studied in \citep{TibshiraniTaylor10GenLasso} drops the last term and consequently excludes inequality constrained applications.

According to the rules of the convex calculus \citep{Ruszczynski06Book}, the unique optimal point $\bx(\rho)$ of the function ${\cal E}_\rho(\bx)$ is characterized by the stationarity condition
\begin{eqnarray}
{\bf 0} & = & \bA \bx(\rho) + \bb + \rho \sum_{i=1}^r s_i \bv_i + \rho \sum_{j=1}^s t_j \bw_j \label{path_stationary}
\end{eqnarray}
with coefficients
\begin{align}
   s_i \in \begin{cases}
   \{-1\} & \bv_i^t \bx - d_i < 0 \\
   [-1, 1] & \bv_i^t \bx - d_i = 0    \\
   \{1\} & \bv_i^t \bx - d_i > 0
   \end{cases}, \hspace{.5in} t_j \in \begin{cases}
   \{0\} & \bw_j^t \bx - e_i < 0 \\
   [0, 1] & \bw_j^t \bx - e_i = 0    \\
   \{1\} & \bw_j^t \bx - e_i > 0
   \end{cases}. \label{eqn:path-subgradient}
\end{align}
Assuming the vectors $\left( \cup_i \{\bv_i\}\right) \cup \left(\cup_j \{\bw_j\}\right)$ are linearly independent, the coefficients $s_i$ and $t_j$ are uniquely determined.  The sets defining the possible values of $s_i$ and $t_j$ are the subdifferentials of the functions
$|s_i|$ and $(t_j)_+=\max\{0,t_j\}$.

The solution path $\bx(\rho)$ is continuous when $\bA$ is positive definite. This also implies that the coefficient paths $\bs(\rho)$ and $\bt(\rho)$ are continuous. For a rigorous proof, note that the representation
\begin{eqnarray*}
\bx(\rho) & = & - \bA^{-1}\Big(\bb + \rho \sum_{i=1}^r s_i \bv_i + \rho \sum_{j=1}^s t_j \bw_j  \Big)
\end{eqnarray*}
entails the norm inequality
\begin{eqnarray*}
\|\bx(\rho)\| & \le & \|\bA^{-1}\|\Big(\|\bb\| + \rho \sum_{i=1}^r \|\bv_i\| + \rho \sum_{j=1}^s \|\bw_j\|  \Big) .
\end{eqnarray*}
Thus, the solution vector $\bx(\rho)$ is bounded whenever $\rho  \ge 0$ is bounded above. To prove continuity, suppose that it fails for a given $\rho$.  Then there exists an $\epsilon>0$ and a sequence $\rho_n$ tending to $\rho$ such $\|\bx(\rho_n)-\bx(\rho)\| \ge \epsilon$ for all $n$. Since $\bx(\rho_n)$ is bounded, we can pass to a subsequence if necessary and assume that $\bx(\rho_n)$ converges to some point $\by$.  Taking limits in the inequality ${\cal E}_{\rho_n}[\bx(\rho_n)] \le {\cal E}_{\rho_n}(\bx)$ demonstrates that ${\cal E}_{\rho}(\by) \le {\cal E}_{\rho}(\bx)$ for all $\bx$. Because $\bx(\rho)$ is unique, we reach the contradictory conclusions $\|\by-\bx(\rho)\| \ge \epsilon$ and $\by = \bx(\rho)$.  Continuity is inherited by the coefficients $s_i$ and $t_j$.  Indeed, let $\bV$ and $\bW$ be the matrices with rows $\bv_i^t$ and $\bw_j^t$, and let $\bU$ be the block matrix $ \begin{pmatrix} \bV \\ \bW \end{pmatrix}$.
The stationarity condition can be restated as
\begin{eqnarray*}
{\bf 0} & = & \bA \bx+\bb+\rho \bU^t \begin{pmatrix}\bs \\ \bt\end{pmatrix} .
\end{eqnarray*}
Multiplying this equation by $\bU$ and solving give
\begin{eqnarray}
\rho \begin{pmatrix}\bs \\ \bt\end{pmatrix} & = & -(\bU \bU^t)^{-1}\bU \Big[\bA\bx(\rho)+\bb \Big], \label{active_mulipliers}
\end{eqnarray}
and the continuity of the left-hand side follows from the continuity of $\bx(\rho)$. Finally, dividing by $\rho$ yields the continuity of the coefficient $s_i$ and $t_j$ for $\rho>0$.

We next show that the solution path is piecewise linear. Along the path we keep track of the following index sets determined by the constraint residuals:
\begin{align*}
   {\cal N}_{\text{E}} &= \{i: \bv_i^t \bx - d_i < 0\}, \hspace{.5in} {\cal N}_{\text{I}} = \{j: \bw_j^t \bx - e_j < 0\}    \\
   {\cal Z}_{\text{E}} &= \{i: \bv_i^t \bx - d_i = 0\}, \hspace{.5in} {\cal Z}_{\text{I}} = \{j: \bw_j^t \bx - e_j = 0\}    \\
   {\cal P}_{\text{E}} &= \{i: \bv_i^t \bx - d_i > 0\}, \hspace{.5in} {\cal P}_{\text{I}} = \{j: \bw_j^t \bx - e_j > 0\}.
\end{align*}
For the sake of simplicity, assume that at the beginning of the current segment $s_i$ does not equal $-1$ or $1$ when $i \in {\cal Z}_{\text{E}}$ and $t_j$ does not equal $0$ or $1$ when $j \in {\cal Z}_{\text{I}}$. In other words, the coefficients of the active constraints occur on the interior of their subdifferentials.  Let us show in this circumstance that the solution path can be extended in a linear fashion. The general idea is to impose the equality constraints $\bV_{{\cal Z}_{\text{E}}} \bx = \bd_{{\cal Z}_{\text{E}}}$ and $\bW_{{\cal Z}_{\text{I}}} \bx = \be_{{\cal Z}_{\text{I}}}$ and write the objective function ${\cal E}_{\rho}(\bx)$ as
\begin{align*}
   \frac 12 \bx^t \bA \bx + \bb^t \bx + c - \rho \sum_{i \in {\cal N}_{\text{E}}} (\bv_i^t \bx - d_i) + \rho \sum_{i \in {\cal P}_{\text{E}}} (\bv_i^t \bx - d_i) + \rho \sum_{j \in {\cal P}_{\text{I}}} (\bw_j^t \bx - e_j).
\end{align*}
For notational convenience define
\begin{align*}
& \bU_{{\cal Z}} = \left( \begin{array}{c} \bV_{{\cal Z}_{\text{E}}} \\ \bW_{{\cal Z}_{\text{I}}} \end{array} \right), \hspace{.15in} \bc_{{\cal Z}} = \left( \begin{array}{c} \bd_{{\cal Z}_{\text{E}}} \\ \be_{{\cal Z}_{\text{I}}} \end{array} \right), \hspace{.15in}
   \bu_{\bar{\cal Z}} = - \sum_{i \in {\cal N}_{\text{E}}} \bv_i + \sum_{i \in {\cal P}_{\text{E}}} \bv_i + \sum_{j \in {\cal P}_{\text{I}}} \bw_j.
\end{align*}
Minimizing ${\cal E}_{\rho}(\bx)$ subject to the constraints generates the Lagrange multiplier problem
\begin{eqnarray}
   \left( \begin{array}{ccc}
   \bA & \bU_{{\cal Z}}^t   \\
   \bU_{{\cal Z}} & {\bf 0}
   \end{array} \right) \left( \begin{array}{c}
   \bx   \\
   \boldsymbol{\lambda}_{{\cal Z}}
   \end{array} \right) & = & \left( \begin{array}{c}
   - \bb - \rho \bu_{\bar {\cal Z}}   \\
   \bc_{{\cal Z}}
   \end{array} \right)  \label{eqn:KKT-matrix}
\end{eqnarray}
with the explicit path solution and Lagrange multipliers
\begin{eqnarray}
   \bx(\rho) & = & - \bP (\bb + \rho \bu_{\bar {\cal Z}}) + \bQ \bc_{{\cal Z}} \;\:\, = \;\:\, - \rho \bP \bu_{\bar {\cal Z}} -\bP \bb+\bQ \bc_{{\cal Z}}  \label{eqn:path-solution} \\
   \blambda_{\cal Z} & = & - \bQ^t \bb + \bR \bc_{{\cal Z}} - \rho \bQ^t \bu_{\bar{\cal Z}}. \label{lagrange-multipiers}
\end{eqnarray}
Here
\begin{eqnarray*}
\begin{pmatrix} \bP & \bQ \\
   \bQ^t & \bR \end{pmatrix}
& = &
\begin{pmatrix} \bA & \bU_{{\cal Z}}^t  \\
\bU_{{\cal Z}} & {\bf 0} \end{pmatrix}^{-1}
\end{eqnarray*}
with
\begin{eqnarray*}
\bP & = & \bA^{-1} - \bA^{-1} \bU_{{\cal Z}}^t ( \bU_{{\cal Z}} \bA^{-1} \bU_{{\cal Z}}^t )^{-1} \bU_{{\cal Z}} \bA^{-1}  \\
\bQ & = & \bA^{-1} \bU_{{\cal Z}}^t ( \bU_{{\cal Z}} \bA^{-1} \bU_{{\cal Z}}^t )^{-1} \\
\bR & = & -(\bU_{{\cal Z}} \bA^{-1} \bU_{{\cal Z}}^t)^{-1}.
\end{eqnarray*}
As we will see in the next section, these seemingly complicated objects arise naturally if path following is organized around the sweep operator.

It is clear that as we increase $\rho$, the solution path (\ref{eqn:path-solution}) changes in a linear fashion until either an inactive constraint becomes active or the coefficient of an active constraint hits the boundary of its subdifferential. We investigate the first case first. Imagining $\rho$ to be a time parameter, an inactive constraint $i \in {\cal N}_{\text{E}} \cup {\cal P}_{\text{E}}$ becomes active when
\begin{eqnarray*}
\bv_i^t \bx(\rho) & = & - \bv_i^t \bP (\bb + \rho \bu_{\bar {\cal Z}}) + \bv_i^t \bQ \bc_{{\cal Z}}  \;\:\, = \;\:\, d_i.
\end{eqnarray*}
If this event occurs, it occurs at the hitting time
\begin{eqnarray}
\rho^{(i)} & = & \frac{- \bv_i^t \bP \bb + \bv_i^t \bQ \bc_{{\cal Z}} - d_i}{\bv_i^t \bP \bu_{\bar {\cal Z}}}. \label{hitting_time1}
\end{eqnarray}
Similarly, an inactive constraint $j \in {\cal N}_{\text{I}} \cup {\cal P}_{\text{I}}$ becomes active at the hitting time
\begin{eqnarray}
\rho^{(j)} & = & \frac{- \bw_j^t \bP \bb + \bw_j^t \bQ \bc_{{\cal Z}} - e_j}{\bw_j^t \bP \bu_{\bar {\cal Z}}}. \label{hitting_time2}
\end{eqnarray}

To determine the escape time for an active constraint, consider once again the stationarity condition (\ref{path_stationary}). The Lagrange multiplier corresponding to an active constraint coincides with a product $\rho s_i(\rho)$ or $\rho t_j(\rho)$. Therefore, if we collect the coefficients for the active constraints into the vector $\br_{{\cal Z}}(\rho)$, then equation (\ref{lagrange-multipiers}) implies
\begin{eqnarray}
\br_{{\cal Z}}(\rho) & = & \frac{1}{\rho} \blambda_{\cal Z}(\rho) \;\:\, = \;\:\,
\frac{1}{\rho} (- \bQ^t \bb + \bR \bc_{{\cal Z}}) - \bQ^t \bu_{\bar{\cal Z}}. \label{rb_solution}
\end{eqnarray}
Formula (\ref{rb_solution}) for $\br_{{\cal Z}}(\rho)$ can be rewritten in terms of the value $\br_{\cal Z}(\rho_0)$ at the start $\rho_0$ of the current segment as
\begin{eqnarray}
    \br_{{\cal Z}}(\rho) &=& \frac{\rho_0}{\rho} \br_{{\cal Z}}(\rho_0) - \left( 1 - \frac{\rho_0}{\rho} \right)\bQ^t \bu_{\bar {\cal Z}}.    \label{eqn:boundary-coeff}
\end{eqnarray}
It is clear that $\br_{{\cal Z}}(\rho)_i$ is increasing in $\rho$ when $[\br_{{\cal Z}}(\rho_0)+\bQ^t\bu_{\bar {\cal Z}}]_i < 0$ and decreasing in $\rho$ when the reverse is true. The coefficient of an active constraint $i \in {\cal Z}_{\text{E}}$ escapes at either of the times
\begin{eqnarray*}
\rho^{(i)} & = & \frac{[-\bQ^t \bb + \bR \bc_{\cal Z}]_i}{[\bQ^t \bu_{\bar {\cal Z}}]_i - 1} \;\; \text{ or } \;\; \frac{[-\bQ^t \bb + \bR \bc_{\cal Z}]_i}{[\bQ^t \bu_{\bar {\cal Z}}]_i + 1},
\end{eqnarray*}
whichever is pertinent.  Similarly, the coefficient of an active constraint $j \in {\cal Z}_{\text{I}}$ escapes at either of the times
\begin{eqnarray*}
\rho^{(j)} & = & \frac{[-\bQ^t \bb + \bR \bc_{\cal Z}]_j}{[\bQ^t \bu_{\bar {\cal Z}}]_j} \;\; \text{ or } \;\; \frac{[-\bQ^t \bb + \bR \bc_{\cal Z}]_j}{[\bQ^t \bu_{\bar {\cal Z}}]_j + 1},
\end{eqnarray*}
whichever is pertinent. The earliest hitting time or escape time over all constraints determines the duration of the current linear segment.

At the end of the current segment, our assumption that all active coefficients occur on the interior of their subdifferentials is actually violated. When the hitting time for an inactive constraint occurs first, we move the constraint to the appropriate active set ${\cal Z}_{\text{E}}$ or ${\cal Z}_{\text{I}}$ and keep the other constraints in place. Similarly, when the escape time for an active constraint occurs first, we move the constraint to the appropriate inactive set and keep the other constraints in place.  In the second scenario, if $s_i$ hits the value $-1$, then we move $i$ to ${\cal N}_{\text{E}}$. If $s_i$ hits the value $1$, then we move $i$ to ${\cal P}_{\text{E}}$.  Similar comments apply when a coefficient $t_j$ hits 0 or 1. Once this move is executed, we commence a new linear segment as just described. The path following algorithm continues segment by segment until for sufficiently large $\rho$ the sets ${\cal N}_{\text{E}}$, ${\cal P}_{\text{E}}$, and ${\cal P}_{\text{I}}$ are exhausted, $\bu_{\bar {\cal Z}} = {\bf 0}$,  and the solution vector (\ref{eqn:path-solution}) stabilizes.

This description omits two details. First, to get the process started, we set $\rho=0$ and $\bx(0)= - \bA^{-1}\bb$. In other words, we start at the unconstrained minimum. For inactive constraints, the coefficients $s_i(0)$ and $t_j(0)$ are fixed. However for active constraints, it is unclear how to assign the coefficients and whether to release the constraints from active status as $\rho$ increases.   Second, very rarely some of the hitting times and escape times will coincide. We are then faced again with the problem of which of the active constraints with coefficients on their subdifferential boundaries to keep active and which to encourage to go inactive in the next segment. In practice, the first problem can easily occur. Roundoff error typically keeps the second problem at bay.

In both anomalous cases, the status of each of active constraint can be resolved by trying all possibilities. Consider the second case first. If there are $a$ currently active constraints parked at their subdifferential boundaries, then there are $2^{a}$ possible configurations for their active-inactive states in the next segment. For a given configuration, we can exploit formula (\ref{rb_solution}) to check whether the coefficient for an active constraint occurs in its subdifferential.  If the coefficient occurs on the boundary of its subdifferential, then we can use representation (\ref{eqn:boundary-coeff}) to check whether it is headed into the interior of the subdifferential as $\rho$ increases.  Since the path and its coefficients are unique, one and only one configuration should determine the next linear segment.  At the start of the path algorithm, the correct configuration also determines the initial values of the active coefficients.  If we take limits in equation (\ref{rb_solution}) as $\rho$ tends to 0, then the coefficients will escape their subdifferentials unless $-\bQ^t \bb + \bR \bc_{\cal Z} = {\bf 0}$ and all components of $-\bQ^t \bu_{\bar {\cal Z}}$ lie in their appropriate subdifferentials.  Hence, again it is easy to decide on the active set ${\cal Z}$ going forward from $\rho=0$.  One could object that the number of configurations $2^a$ is potentially very large, but in practice this combinatorial bottleneck never occurs.  Visiting the various configurations can be viewed as a systematic walk through the subsets of $\{1,\ldots,a\}$ and organized using a classical gray code  \citep{Savage97GrayCodes} that deletes at most one element and adjoins at most one element as one passes from one active subset to the next.  As we will see in the next section, adjoining an element corresponds to sweeping a diagonal entry of a tableau and deleting an element corresponds to inverse sweeping a diagonal entry of the same tableau.

\section{The Path Algorithm and Sweeping}
\label{sec:implementation}

Implementation of the path algorithm can be conveniently organized around the sweep and inverse sweep operators of regression analysis \citep{Dempster69Book,Goodnight79Sweep,Jennrich77Stepwisereg,LittleRubin02Book,Lange10NumAnalBook}. We first recall the definition and basic properties of the sweep operator. Suppose $\bA$ is an $m \times m$ symmetric matrix. Sweeping on the $k$th diagonal entry
$a_{kk} \ne 0$ of $\bA$ yields a new symmetric matrix $\widehat{\bA}$ with entries
\begin{eqnarray*}
    \hat{a}_{kk} &=& - \frac{1}{a_{kk}}, \\
    \hat{a}_{ik} &=& \frac{a_{ik}}{a_{kk}}, \quad i \ne k  \\
    \hat{a}_{kj} &=& \frac{a_{kj}}{a_{kk}}, \quad j \ne k   \\
    \hat{a}_{ij} &=& a_{ij} - \frac{a_{ik}a_{kj}}{a_{kk}}, \quad i,j \ne k .
\end{eqnarray*}
These arithmetic operations can be undone by inverse sweeping on the same diagonal entry. Inverse sweeping sends the symmetric matrix
$\bA$ into the symmetric matrix $\check{\bA}$ with entries
\begin{eqnarray*}
    \check{a}_{kk} &=& - \frac{1}{a_{kk}}, \\
    \check{a}_{ik} &=&  - \frac{a_{ik}}{a_{kk}}, \quad i \ne  k\\
    \check{a}_{kj} &=& - \frac{a_{kj}}{a_{kk}},  \quad j \ne k\\
    \check{a}_{ij} &=& a_{ij} - \frac{a_{ik}a_{kj}}{a_{kk}}, \quad i,j \ne k.
\end{eqnarray*}
Both sweeping and inverse sweeping preserve symmetry. Thus, all operations can be carried out on either the lower or upper triangle of
$\bA$ alone, saving both computational time and storage. When several sweeps or inverse sweeps are performed, their order is irrelevant.
Finally, a symmetric matrix $\bA$ is positive definite if and only if $\bA$ can be completely swept, and all of its diagonal entries remain positive until swept.  Complete sweeping produces $-\bA^{-1}$.  Each sweep of a positive definite matrix reduces the magnitude of the
unswept diagonal entries. Positive definite matrices with poor condition numbers can be detected by monitoring the relative magnitude of
each diagonal entry just prior to sweeping.

At the start of path following, we initialize a path tableau with block entries
\begin{eqnarray}
   \left( \begin{array}{c|cc}
   -\bA & -\bU^t & \bb \\
   \hline
   * & {\bf 0} & -\bc  \\
   * & * & 0
   \end{array} \right). \label{eqn:initial-tableau}
\end{eqnarray}
The starred blocks here are determined by symmetry. Sweeping the diagonal entries of the upper-left block $-\bA$ of the tableau yields
\begin{eqnarray*}
   \left( \begin{array}{c|cc}
   \bA^{-1} & \bA^{-1} \bU^t & - \bA^{-1} \bb \\
   \hline
   * & \bU \bA^{-1} \bU^t & - \bU \bA^{-1} \bb - \bc  \\
   * & * & \bb^t \bA^{-1} \bb
   \end{array} \right). 
\end{eqnarray*}
The new tableau contains the unconstrained solution $\bx(0) = - \bA^{-1} \bb$ and the corresponding constraint residuals $- \bU \bA^{-1} \bb - \bc$. In path following, we adopt our previous notation and divide the original tableau into sub-blocks. The result
\begin{eqnarray}
   \left( \begin{array}{cc|cc}
   -\bA & -\bU_{{\cal Z}}^t & -\bU_{\bar{\cal Z}}^t & \bb \\
   * & {\bf 0} & {\bf 0} & -\bc_{{\cal Z}}  \\
   \hline
   * & * & {\bf 0} & -\bc_{\bar {\cal Z}}   \\
   * & * & * & 0
   \end{array} \right)  \label{eqn:path-tableau}
\end{eqnarray}
highlights the active and inactive constraints. If we continue sweeping until all diagonal entries of the upper-left
quadrant of this version of the tableau are swept, then the tableau becomes
\begin{eqnarray*}
   \left( \begin{array}{cc|cc}
   \bP & \bQ & \bP \bU_{\bar{\cal Z}}^t & -\bP \bb + \bQ \bc_{{\cal Z}} \\
   * & \bR & \bQ^t \bU_{\bar{\cal Z}}^t & -\bQ^t \bb + \bR \bc_{{\cal Z}} \\
   \hline
   * & * & \bU_{\bar{\cal Z}} \bP \bU_{\bar {\cal Z}}^t & \bU_{\bar{\cal Z}}(-\bP\bb+\bQ\bc_{{\cal Z}})-\bc_{\bar{\cal Z}}   \\
   * & * & * & \bb^t \bP \bb - 2 \bb^t \bQ \bc_{{\cal Z}} + \bc_{{\cal Z}}^t \bR \bc_{{\cal Z}}
   \end{array} \right).
\end{eqnarray*}
All of the required elements for the path algorithm now magically appear.

Given the next $\rho$, the solution vector $\bx(\rho)$ appearing in equation (\ref{eqn:path-solution}) requires the sum
$-\bP\bb+\bQ\bc_{{\cal Z}}$, which occurs in the revised tableau, and the vector $\bP \bu_{\bar {\cal Z}}$. If $\br_{\bar {\cal Z}}$ denotes the coefficient vector for the inactive constraints, with entries of $-1$ for constraints in ${\cal N}_{\text{E}}$, 0 for constraints in
${\cal N}_{\text{I}}$, and 1 for constraints in ${\cal P}_{\text{E}} \cup {\cal P}_{\text{I}}$, then
$\bP \bu_{\bar {\cal Z}} = \bP \bU_{\bar{\cal Z}}^t \br_{\bar {\cal Z}}$. Fortunately, $\bP \bU_{\bar{\cal Z}}^t$ appears in the revised tableau. The update of $\rho$ depends on the hitting times (\ref{hitting_time1}) and (\ref{hitting_time2}). These in turn depend on the numerators $- \bv_i^t \bP \bb + \bv_i^t \bQ \bc_{{\cal Z}} - d_i$ and $- \bw_j^t \bP \bb + \bw_j^t \bQ \bc_{{\cal Z}} - e_j$, which occur as components of the vector  $\bU_{\bar{\cal Z}}(-\bP\bb+\bQ\bc_{{\cal Z}})-\bc_{\bar{\cal Z}}$, and the denominators $\bv_i^t \bP \bu_{\bar {\cal Z}}$ and $\bw_j^t \bP \bu_{\bar {\cal Z}}$, which occur as components of the matrix
$\bU_{\bar{\cal Z}} \bP \bU_{\bar {\cal Z}}^t \br_{\bar {\cal Z}}$ computable from the block $\bU_{\bar{\cal Z}} \bP \bU_{\bar {\cal Z}}^t $
of the tableau. The escape times for the active constraints also determine the update of $\rho$.  According to equation
(\ref{eqn:boundary-coeff}), the escape times depend on the current coefficient vector, the current value $\rho_0$
of $\rho$, and the vector $\bQ^t \bu_{\bar {\cal Z}} = \bQ^t \bU_{\bar{\cal Z}}^t \br_{\bar {\cal Z}}$, which can be computed from the
block $ \bQ^t \bU_{\bar{\cal Z}}^t$ of the tableau.  Thus, the revised tableau supplies all of the ingredients for path following.  Algorithm \ref{algo:primal-path} outlines the steps for path following ignoring the anomalous situations.

The ingredients for handling the anomalous situations can also be read from the path tableau.  The initial coefficients $\br_{{\cal Z}}(0) = - \bQ^t \bu_{\bar {\cal Z}} = \bQ^t \bU_{\bar{\cal Z}}^t \br_{\bar {\cal Z}}$ are available once we sweep the tableau (\ref{eqn:initial-tableau}) on the diagonal entries corresponding to  the constraints in ${\cal Z}$ at the point $\bx(0) = - \bA^{-1} \bb$.  As noted earlier, if the coefficients of several active constraints are simultaneously poised to exit their subdifferentials, then one must consider all possible swept and unswept combinations of these constraints.  The operative criteria for choosing the right combination involve the available quantities $\bQ^t \bu_{\bar {\cal Z}}$ and $-\bQ^t \bb + \bR \bc_{{\cal Z}}$. One of the sweeping combinations is bound to give a correct direction for the next extension of the path.

The computational complexity of path following depends on the number of parameters $m$ and the number of constraints $n=r+s$. Computation of the initial solution $-\bA^{-1} \bb$ takes about $3 m^3$ floating point operations (flops). There is no need to store or update the
$\bP$ block during path following. The remaining sweeps and inverse sweeps take on the order of $n(m+n)$ flops each.  This count must be multiplied by the number of segments along the path, which empirically is on the order of $O(n)$. The sweep tableau requires storing
$(m+n)^2$ real numbers.  We recommend all computations be done in double precision. Both flop counts and storage can be halved by exploiting symmetry. Finally, it is worth mentioning some computational shortcuts for the multi-task learning model. Among these are
the formulas
\begin{eqnarray*}
(\bI \otimes \bX)^t (\bI \otimes \bX) & = & \bI \otimes \bX^t \bX \\
(\bI \otimes \bX^t \bX)^{-1} & = & \bI \otimes (\bX^t \bX)^{-1} \\
(\bI \otimes \bX^t \bX)^{-1} \bI \otimes \bV & = & \bI \otimes (\bX^t \bX)^{-1}\bV \\
(\bI \otimes \bX^t \bX)^{-1} \bI \otimes \bW & = & \bI \otimes (\bX^t \bX)^{-1}\bW .
\end{eqnarray*}

\begin{algorithm}
\begin{algorithmic}
\STATE Initialize $k=0$, $\rho_0 = 0$, and the path tableau (\ref{eqn:initial-tableau}). Sweep the diagonal entries of $-\bA$. Enter the main loop.
\REPEAT
\STATE Increment $k$ by 1.
\STATE Compute the hitting time or exit time $\rho^{(i)}$ for each constraint $i$.
\STATE Set $\rho_k = \min \{\rho^{(i)}: \rho^{(i)} > \rho_{k-1}\}$.
\STATE Update the coefficient vector by equation (\ref{eqn:boundary-coeff}).
\STATE Sweep the diagonal entry of the inactive constraint that becomes active or inverse sweep the diagonal entry of the active constraint that becomes inactive.
\STATE Update the solution vector $\bx_k =\bx(\rho_{k})$ by equation (\ref{eqn:path-solution}).
\UNTIL{${\cal N}_{\text{E}} = {\cal P}_{\text{E}} = {\cal P}_{\text{I}} = \emptyset$}
\end{algorithmic}
\caption{Solution path of the primal problem (\ref{eqn:path-penalized-obj}) when $\bA$ is positive definite.}
\label{algo:primal-path}
\end{algorithm}

\section{Extensions of the Path Algorithm \label{path_extensions_section}}

As just presented, the path algorithm starts from the unconstrained solution and moves forward along the path to the constrained solution. With minor modifications, the same algorithm can start in the middle of the path or move in the reverse direction along it.  The latter tactic might prove useful in lasso and fused-lasso problems, where the fully constrained solution is trivial. In general, consider starting from $\bx(\rho_0)$ at a point $\rho_0$ on the path. Let ${\cal Z} = {\cal Z}_{\text{E}} \cup {\cal Z}_{\text{I}}$ continue to denote the zero set for the segment containing $\rho_0$.  Path following begins by sweeping the upper left block of the tableau (\ref{eqn:path-tableau}) and then proceeds as indicated in Algorithm \ref{algo:primal-path}. Traveling in the reverse direction entails calculation of hitting and exit times for decreasing $\rho$ rather than increasing $\rho$.

Our assumption that $\bA$ is positive definite automatically excludes underdetermined statistical problems with more parameters than cases.  Here we briefly indicate how to carry out the exact penalty method when this assumption fails and the sweep operator cannot be brought into play.  In the absence of constraints, $f(\bx)$ lacks a minimum if and only if either $\bA$ has a negative eigenvalue or the equation $\bA \bx =  \bb$ has no solution.  In either circumstance a unique global minimum may exist if enough constraints are enforced. Suppose $\bx(\rho_0)$ supplies the minimum of the exact penalty function ${\cal E}_{\rho}(\bx)$ at $\rho=\rho_0>0$. Let the matrix $\bU_{\cal Z}$ hold the active constraint vectors.  As we slide along the active constraints, the minimum point can be represented as $\bx(\rho) = \bx(\rho_0)+ \bY \by(\rho)$, where the columns of $\bY$ are orthogonal to the rows of $\bU_{\cal Z}$. One can construct $\bY$ by the Gramm-Schmidt process; $\bY$ is then the orthogonal complement of $\bU_{\cal Z}$ in the QR decomposition. The active constraints hold because $\bU_{\cal Z} \bx(\rho) = \bU_{\cal Z} \bx(\rho_0) = \bc_{{\cal Z}}$.

The analogue of the stationarity condition (\ref{path_stationary}) under reparameterization is
\begin{eqnarray}
{\bf 0} & = & \bY^t \bA \bY \by(\rho) + \bY^t \bb + \rho  \bY^t  \bu_{\bar{\cal Z}} \label{path_stationary2}.
\end{eqnarray}
The inactive constraints do not appear in this equation because $\bv_i^t \bY = {\bf 0}$ and $\bw_j^t \bY = {\bf 0}$ for $i$ or $j$
active.  Solving for $\by(\rho)$ and $\bx(\rho)$ gives
\begin{eqnarray}
\by(\rho) & = & - (\bY^t \bA \bY )^{-1}( \bY^t \bb + \rho  \bY^t  \bu_{\bar{\cal Z}}) \nonumber \\
\bx(\rho) & = & \bx(\rho_0)- \bY (\bY^t \bA \bY )^{-1}( \bY^t \bb + \rho  \bY^t  \bu_{\bar{\cal Z}})    \label{eqn:alt-path-sol}
\end{eqnarray}
and does not require inverting $\bA$. Because the solution $\bx(\rho)$ is affine in $\rho$, it is straightforward to calculate the
hitting times for the inactive constraints.

Under the original parametrization, the Lagrange multipliers and corresponding active coefficients appearing in the stationarity condition (\ref{path_stationary}) can still be recovered by invoking equation (\ref{active_mulipliers}). Again it is a simple matter
to calculate exit times.  The formulas are not quite as elegant as those based on the sweep operator, but all essential elements for traversing the path are available.  Adding or deleting a row of the matrix $\bU_{\cal Z}$ can be accomplished by updating the QR decomposition. The fast algorithms for this purpose simultaneously update $\bY$ \citep{LawsonHanson87LSBook,NocedalWright06Book}.
More generally for equality constrained problems generated by the lasso and generalized lasso, the constraint matrix $\bU_{\cal Z}$ as
one approaches the penalized solution is often very sparse. The required QR updates are then numerically cheap.  For the sake of brevity, we omit further details.

\section{Degrees of Freedom Under Affine Constraints}
\label{sec:dof}

We now specialize to the least squares problem with the choices $\bA = \bX^t \bX$, $\bb = -\bX^t \by$, and $\bx(\rho)=\hat \bbeta(\rho)$ and consider how to define degrees of freedom in the presence of both equality and inequality constraints. As previous authors
\citep{EfronHastieIainTibshirani04LARS,TibshiraniTaylor10GenLasso,ZouHastieTibshirani07LassoDF} have shown, the most productive approach relies on Stein's characterization \citep{Efron04CV,Stein81MVN}
\begin{eqnarray*}
 \text{df}(\hat \by) & = &  \mathbf{E} \left(\sum_{i=1}^n \frac{\partial }{\partial y_i} \hat{y_i} \right)
 \;\:\, = \;\:\,\mathbf{E} \left[ \text{tr} ( d_{\by} \hat{\by} )\right]
\end{eqnarray*}
of the degrees of freedom.  Here $\hat{\by} = \bX \hat \bbeta$ is the fitted value of $\by$, and $d_{\by} \hat{\by}$ denotes its differential with respect to
the entries of $\by$. Equation (\ref{eqn:path-solution}) implies that
\begin{eqnarray*}
\hat{\by} & = & \bX \hat{\bbeta} \;\:\, = \;\:\, \bX \bP \bX^t \by + \bX \bQ \bc_{\cal Z} - \rho \bX \bP \bu_{\bar {\cal Z}}.
\end{eqnarray*}
Because $\rho$ is fixed, it follows that $d_{\by} \hat{\by} = \bX \bP \bX^t$.  The representation
\begin{eqnarray*}
    & & \bX \bP \bX^t   \\
    &=& \bX (\bX^t\bX)^{-1} \bX^t - \bX(\bX^t\bX)^{-1} \bU_{\cal Z}^t [\bU_{\cal Z}(\bX^t\bX)^{-1}\bU_{\cal Z}^t]^{-1} \bU_{\cal Z}(\bX^t\bX)^{-1} \bX^t   \\
    &=& \bP_1 - \bP_2
\end{eqnarray*}
and the cyclic permutation property of the trace function applied to the matrices $\bP_1$ and $\bP_2$ yield the formula
\begin{eqnarray*}
\mathbf{E} \left[ \text{tr} ( d_{\by} \hat{\by} )\right]    &=& m - \mathbf{E} ( |{\cal Z}|) ,
\end{eqnarray*}
where $m$ equals the number of parameters. In other words,  $m - |{\cal Z}|$ is an unbiased estimator of the degrees of freedom.  This result obviously depends on our assumptions that $\bX$ has full column rank $m$ and the constraints $\bv_i$ and $\bw_j$ are linearly independent. The latter condition is obviously true for lasso and fused-lasso problems. The validity of Stein's formula requires the fitted value $\hat \by$ to be a continuous and almost surely differentiable function of $\by$ \citep{Stein81MVN}. Fortunately, this is the case for lasso \citep{ZouHastieTibshirani07LassoDF} and generalized lasso problems \citep{TibshiraniTaylor10GenLasso} and for at least one case of shape-restricted regression \citep{MeyersWoodroofe00DF}. Our derivation does not depend directly on whether the constraints are equality or inequality constraints.  Hence, the degrees of freedom estimator can be applied in shape-restricted regression using model selection criteria such as $C_p$, AIC, and BIC along the whole path. The concave regression example in the introduction illustrates the general idea.

\section{Examples}
\label{sec:examples}

Our examples illustrate both the mechanics and the potential of path following. The path algorithm's ability to handle inequality constraints allows us to obtain path solutions to a variety of shape-restricted regressions. Problems of this sort may well dominate the future agenda of nonparametric estimation.

\subsection{Two Toy Examples}

\begin{figure}
{\centering
\includegraphics[width=2.5in]{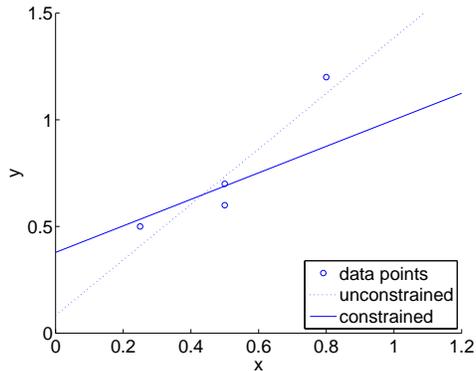}
\caption{The data points and the fitted lines for the first toy example of constrained curve fitting \citep{LawsonHanson87LSBook}.}
\label{fig:toy-lawsonhanson}
}
\end{figure}

Our first example \citep{LawsonHanson87LSBook} fits a straight line $y= \beta_0 + x \beta_1$ to the data points (0.25,0.5), (0.5,0.6), (0.5,0.7), and (0.8,1.2)
by minimizing the least squares criterion $\|\by - \bX \bbeta\|_2^2$ subject to the constraints
\begin{eqnarray*}
   \beta_1 \ge 0, \hspace{.2in} \beta_0 \ge 0, \hspace{.2in} \beta_0+\beta_1 \le 1.
\end{eqnarray*}
In our notation
\begin{eqnarray*}
    \bA &=& \bX^t \bX \;\; = \;\; \left( \begin{array}{cc} 4.0000 & 2.0500 \\ 2.0500 & 1.2025 \end{array} \right), \hspace{.1in} \bb \;\; = \;\;- \bX^t \by \;\; = \;\;\left( \begin{array}{c} -3.0000 \\ -1.7350 \end{array} \right), \\
    \bW &=& \left( \begin{array}{rr} -1 & 0 \\ -1 & 0 \\ 1 & 1 \end{array} \right), \hspace{.1in} \be \;\;= \;\; \left( \begin{array}{c} 0 \\ 0 \\ 1 \end{array} \right).
\end{eqnarray*}
The initial tableau is
\begin{eqnarray*}
\left( \begin{tabular}{rr|rrr|r}
-4.0000&-2.0500&1&1&-1&-3.0000\\
-2.0500&-1.2025&0&0&-1&-1.7350\\
\hline
1&0&0&0&0&0\\
1&0&0&0&0&0\\
-1&-1&0&0&0&-1  \\
\hline
-3.0000&-1.7350&0&0&-1&0
\end{tabular} \right).
\end{eqnarray*}
Sweeping the first two diagonal entries produces
\begin{eqnarray*}
\left( \begin{tabular}{rr|rrr|r}
1.9794&-3.3745&-1.9794&3.3745&-1.3951&0.0835\\
-3.3745&6.5844&3.3745&-6.5844&3.2099&1.3004\\\hline
-1.9794&3.3745&1.9794&-3.3745&1.3951&-0.0835\\
3.3745&-6.5844&-3.3745&6.5844&-3.2099&-1.3004\\
-1.3951&3.2099&1.3951&-3.2099&1.8148&0.3840\\\hline
0.0835&1.3004&-0.0835&-1.3004&0.3840&2.5068\\
\end{tabular}
 \right) ,
\end{eqnarray*}
from which we read off the unconstrained solution $\bbeta(0) =(0.0835,1.3004)^t$ and the constraint residuals $(-0.0835,-1.3004,0.3840)^t$. The latter indicates that ${\cal N}_{\text{I}} = \{1,2\}$, ${\cal Z}_{\text{I}} = \emptyset$, and ${\cal P}_{\text{I}} = \{3\}$. Multiplying the middle block matrix by the coefficient vector $\br = (0, 0, 1)^t$ and dividing the residual vector entrywise give the hitting times $\rho = (-0.0599, 0.4051, 0.2116)$. Thus $\rho_{1} = 0.2116$ and
\begin{eqnarray*}
    \bbeta(0.2116) & = & \left( \begin{array}{c}  0.0835 \\ 1.3004 \end{array} \right) - 0.2116 \times \left( \begin{array}{c} -1.3951 \\ 3.2099 \end{array} \right) \;\:\, = \;\:\, \left( \begin{array}{c} 0.3787 \\ 0.6213 \end{array} \right).
\end{eqnarray*}
Now ${\cal N}=\{1,2\}$, ${\cal Z}=\{3\}$, ${\cal P}=\emptyset$, and we have found the solution. Figure \ref{fig:toy-lawsonhanson} displays the data points and the unconstrained and constrained fitted lines.

\begin{figure}[b]
$$
\begin{array}{cc}
\includegraphics[width=2.5in]{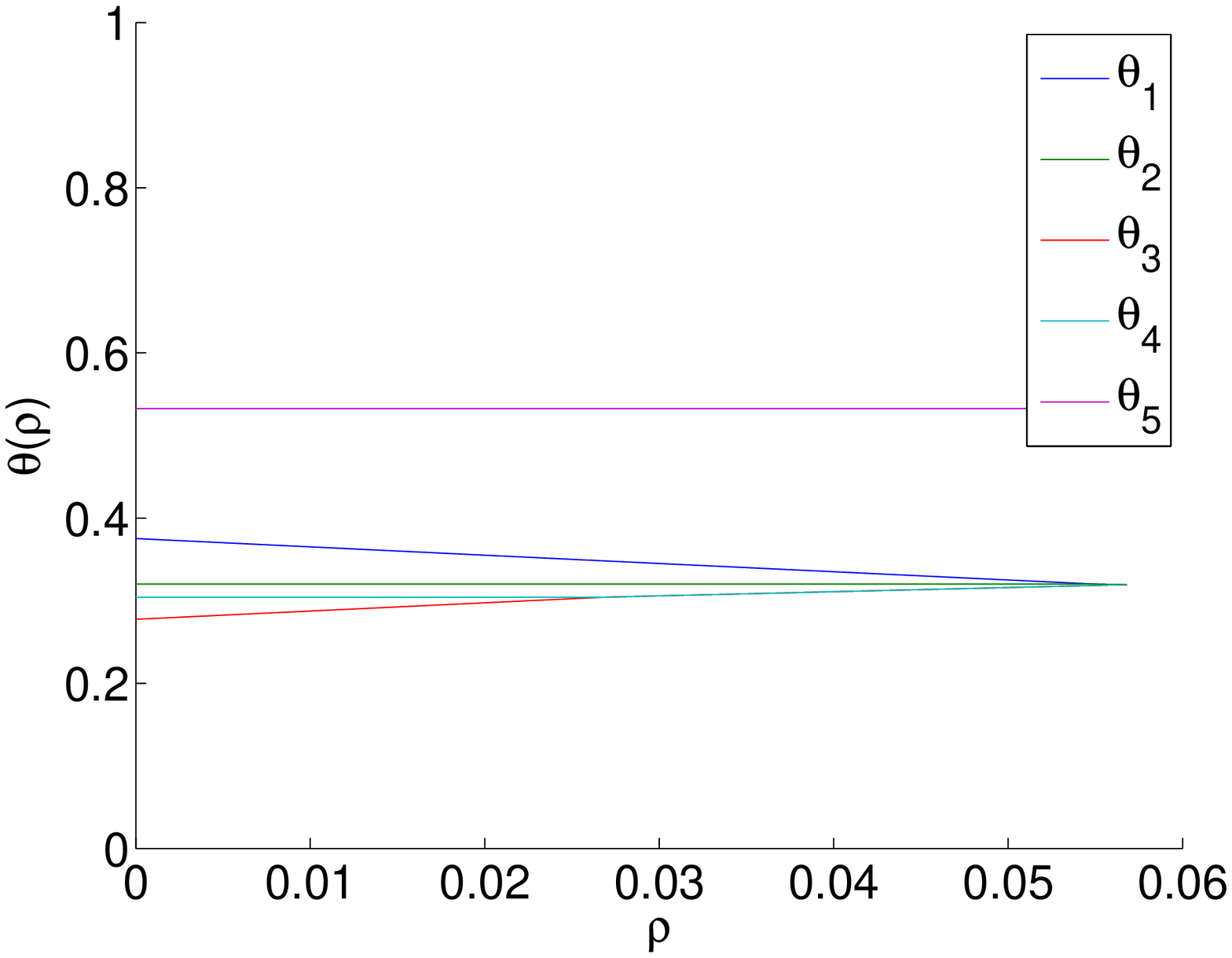} & \includegraphics[width=2.25in]{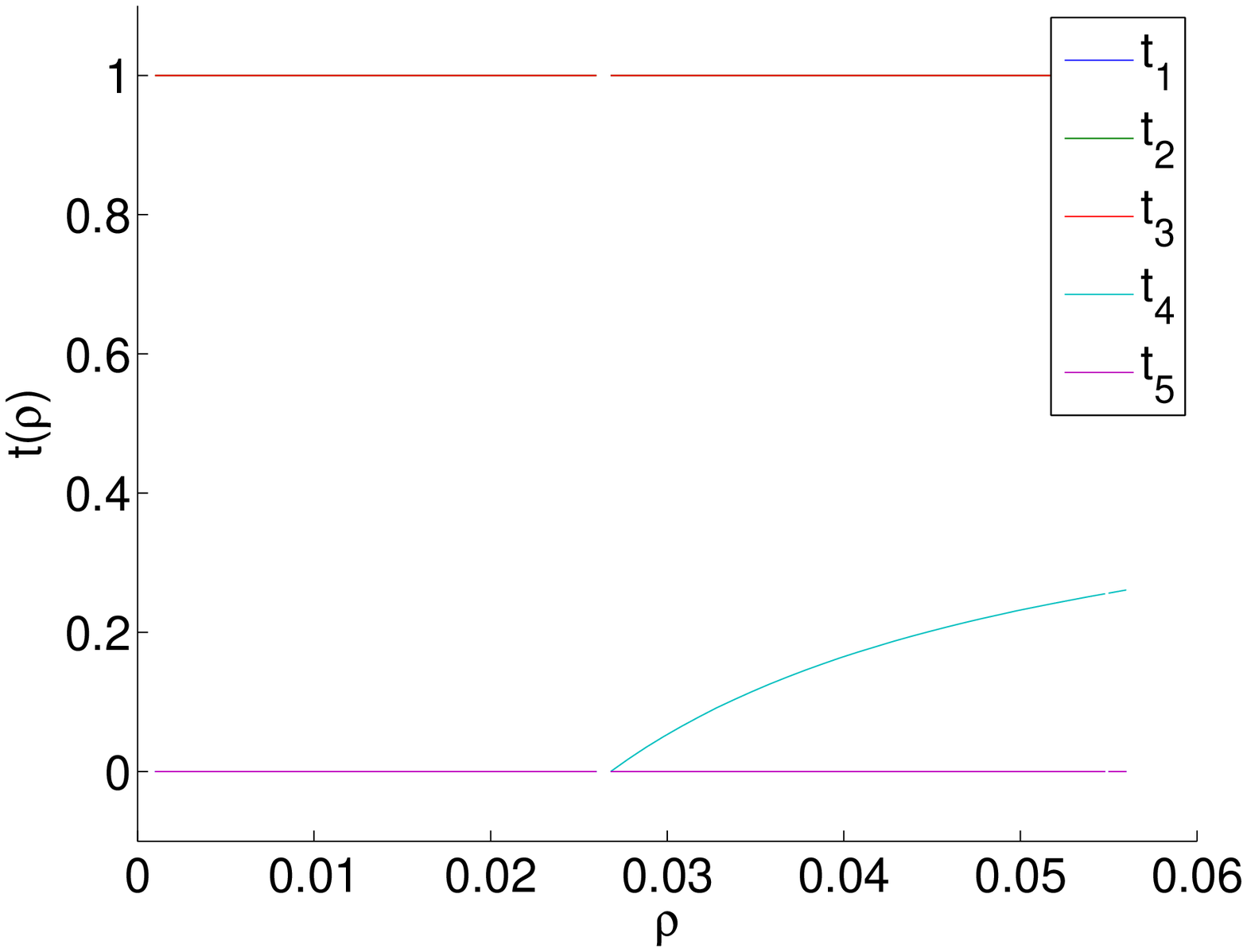}
\end{array}
$$
\caption{Toxin response example. Left: Solution path. Right: Coefficient paths for the constraints.}
\label{fig:solpath-fish}
\end{figure}

Our second toy example concerns the toxin response problem \citep{Schoenfeld86Order} with $m$ toxin levels $x_1 \le x_2 \le \cdots \le x_m$ and a mortality rate $y_i=f(x_i)$ at each level. It is reasonable to assume that the mortality function $f(x)$ is nonnegative and increasing. Suppose $\bar y_i$ are the observed death frequencies averaged across $n_i$ trials at level $x_i$. In a finite sample, the $\bar y_i$ may fail to be nondecreasing. For example, in an EPA study of the effects of chromium on fish \citep{Schoenfeld86Order}, the observed binomial frequencies and chromium levels are
\begin{eqnarray*}
\bar \by & = & (0.3752, 0.3202, 0.2775, 0.3043, 0.5327)^t \\
\bx & = & (51, 105, 194, 384, 822)^t \;\; \mbox{in $\mu$g/l}.
\end{eqnarray*}
Isotonic regression minimizes $\sum_{k=1}^m (\bar y_k - \theta_k)^2$ subject to the constraints $0 \le \theta_1 \le \cdots \le \theta_m$ on the binomial parameters $\theta_k = f(x_k)$. The solution path depicted in Figure \ref{fig:solpath-fish} is continuous and piecewise linear as advertised, but the coefficient paths are nonlinear. The first four binomial parameters coalesce in the constrained estimate.

\subsection{Generalized Lasso Problems}

The generalized lasso problems studied in \citep{TibshiraniTaylor10GenLasso} all reduce to minimization of some form of the objective function (\ref{eqn:path-penalized-obj}). To avoid repetition, we omit detailed discussion of this class of problems and simply refer readers interested in applications to lasso or fused-lasso penalized regression, outlier detections, trend filtering, and image restoration to the original article \citep{TibshiraniTaylor10GenLasso}. Here we would like to point out the relevance of the generalized lasso problems to graph-guided penalized regression \citep{ChenLinKimCarbonellXing10ProxGrad}. Suppose each node $i$ of a graph is assigned a regression coefficient $\beta_i$ and a weight $w_i$.  In graph penalized regression, the objective function takes the form
\begin{eqnarray}
    \frac 12 \| \bW(\by - \bX \bbeta)\|_2^2 + \lambda_{\text{G}} \sum_{i \sim j} \left| \frac{\beta_i}{\sqrt{d_i}} - \text{sgn}(r_{ij})\frac{\beta_j}{\sqrt{d_j}} \right| + \lambda_{\text{L}} \sum_j |\beta_j|, \label{graph_objective}
\end{eqnarray}
where the set of neighboring pairs $i \sim j$ define the graph, $d_i$ is the degree of node $i$, and $r_{ij}$ is the correlation coefficient between $i$ and $j$. Under a line graph, the objective function (\ref{graph_objective}) reduces to the fused lasso. In 2-dimensional imaging applications, the graph consists of neighboring pixels in the plane, and minimization of the function (\ref{graph_objective}) is accomplished by total variation algorithms. In MRI images, the graph is defined by neighboring pixels in 3 dimensions. Penalties are introduced in image reconstruction and restoration to enforce smoothness. In microarray analysis, the graph reflects gene networks. Smoothing the $\beta_i$ over the network is motivated by the assumption that the expression levels of related genes should rise and fall in a coordinated fashion. Ridge regularization in graph penalized regression \citep{LiLi08GraphReg} is achieved by changing the objective function to
\begin{eqnarray*}
    \frac 12 \| \bW(\by - \bX \bbeta)\|_2^2 + \lambda_{\text{G}} \sum_{i \sim j} \left( \frac{\beta_i}{\sqrt{d_i}} - \text{sgn}(r_{ij})\frac{\beta_j}{\sqrt{d_j}} \right)^2 + \lambda_{\text{L}} \sum_j |\beta_j|.
\end{eqnarray*}
If one fixes either of the tuning constants in these models, our path algorithm delivers the solution path as a function of the other tuning constant, Alternatively, one can fix the ratio of the two tuning constants. Finally, the extension
\begin{eqnarray*}
    \frac 12 \|\bY - \bX \bB\|_{\text{F}}^2 + \lambda_{\text{G}} \sum_{i \sim j} \sum_{k=1}^K \left| \frac{\beta_{ki}}{\sqrt{d_i}} - \text{sgn}(r_{ij})\frac{\beta_{kj}}{\sqrt{d_j}} \right| + \lambda_{\text{L}} \sum_{k,i} |\beta_{k,i}|
\end{eqnarray*}
of the objective function to multivariate response models is obvious.

In principle, the path algorithm applies to all of these problems provided the design matrix $\bX$ has full column rank. If $\bX$ has reduced rank, then it is advisable to add a small amount of ridge regularization $\epsilon \sum_i \beta_i^2$ to the objective function \citep{TibshiraniTaylor10GenLasso}. Even so, computation of the unpenalized solution may be problematic in high dimensions. Alternatively, path following can be conducted starting from the fully constrained problem as suggested in Section \ref{path_extensions_section}.

\subsection{Shape Restricted Regressions}

Order-constrained regression is now widely accepted as an important modeling tool
\citep{RobertsonWrightDykstra88Book,SilvapullePranab05CSIBook}.   If $\bbeta$ is the parameter vector, monotone regression includes
isotone constraints $\beta_1 \le \beta_2 \le  \cdots \le \beta_m$ or antitone constraints $\beta_1 \ge \beta_2 \ge  \cdots \ge \beta_m$. In partially ordered regression, subsets of the parameters are subject to isotone or antitone constraints.  In other problems it is sensible to impose convex or concave constraints.  If observations are collected at irregularly spaced time points $t_1 \le t_2 \le \cdots \le t_m$, then convexity translates into the constraints
\begin{eqnarray*}
\frac{\beta_{i+2}-\beta_{i+1}}{t_{i+2}-t_{i+1}} & \ge & \frac{\beta_{i+1}-\beta_i}{t_{i+1}-t_i}
\end{eqnarray*}
for $1 \le i \le m-2$. When the time intervals are uniform, these convex constraints become $\beta_{i+2} - \beta_{i+1} \ge \beta_{i+1} - \beta_{i}$. Concavity translates into the opposite set of inequalities.  All of these shape restricted regression problems can be solved by path following.

\begin{figure}
\centering
$$
\begin{array}{cc}
\includegraphics[width=2.25in]{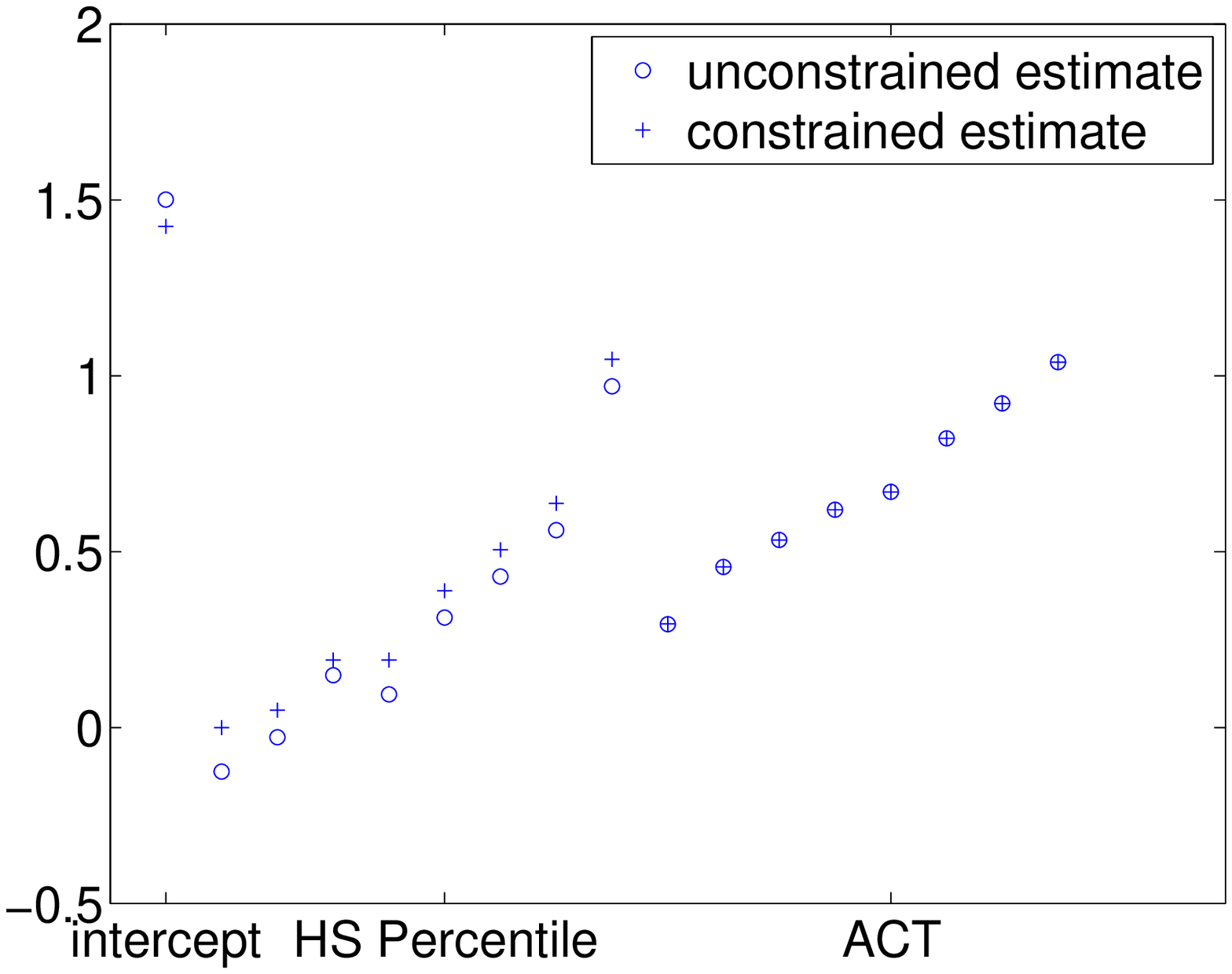} & \includegraphics[width=2.25in]{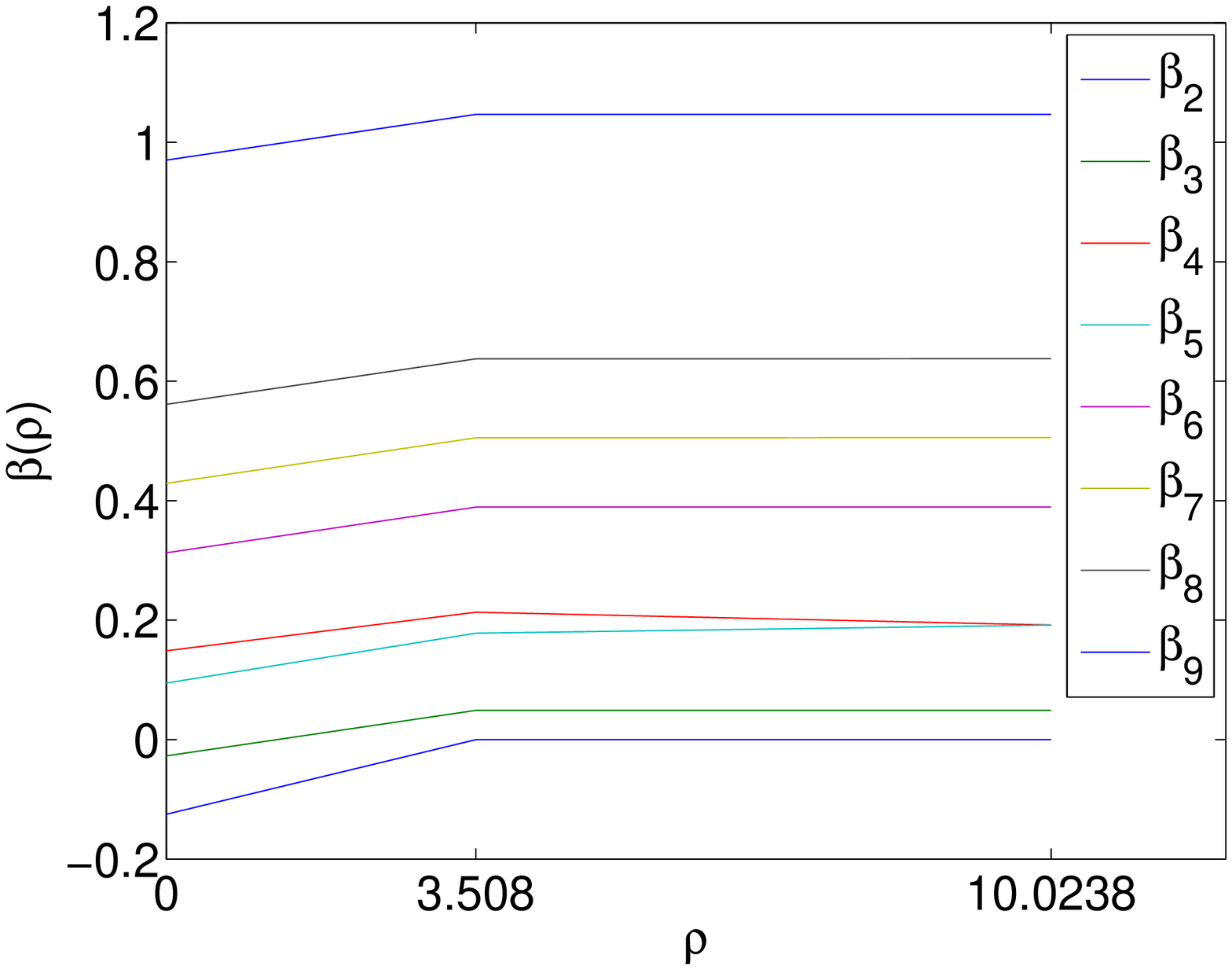}
\end{array}
$$
\caption{Left: Unconstrained and constrained estimates for the Iowa GPA data. Right: Solution paths of for the regression coefficients
corresponding to high school rank.}
\label{fig:iowagpa}
\end{figure}

As an example of partial isotone regression, we fit the data from Table 1.3.1 of the reference \citep{RobertsonWrightDykstra88Book} on
the first-year grade point averages (GPA) of 2397 University of Iowa freshmen.  These data can be downloaded as part of the {\tt R} package {\tt ic.infer}. The ordinal predictors high school rank (as a percentile) and ACT (a standard aptitude test) score are discretized into nine ordered categories each.  A rational admission policy based on these two predictor sets should be isotone separately within each set. Figure \ref{fig:iowagpa} shows the unconstrained and constrained solutions for the intercept and the two predictor sets and the solution path of the regression coefficients for the high school rank predictor.

\begin{figure}
\centering
$$
\begin{array}{cc}
\includegraphics[width=2.25in]{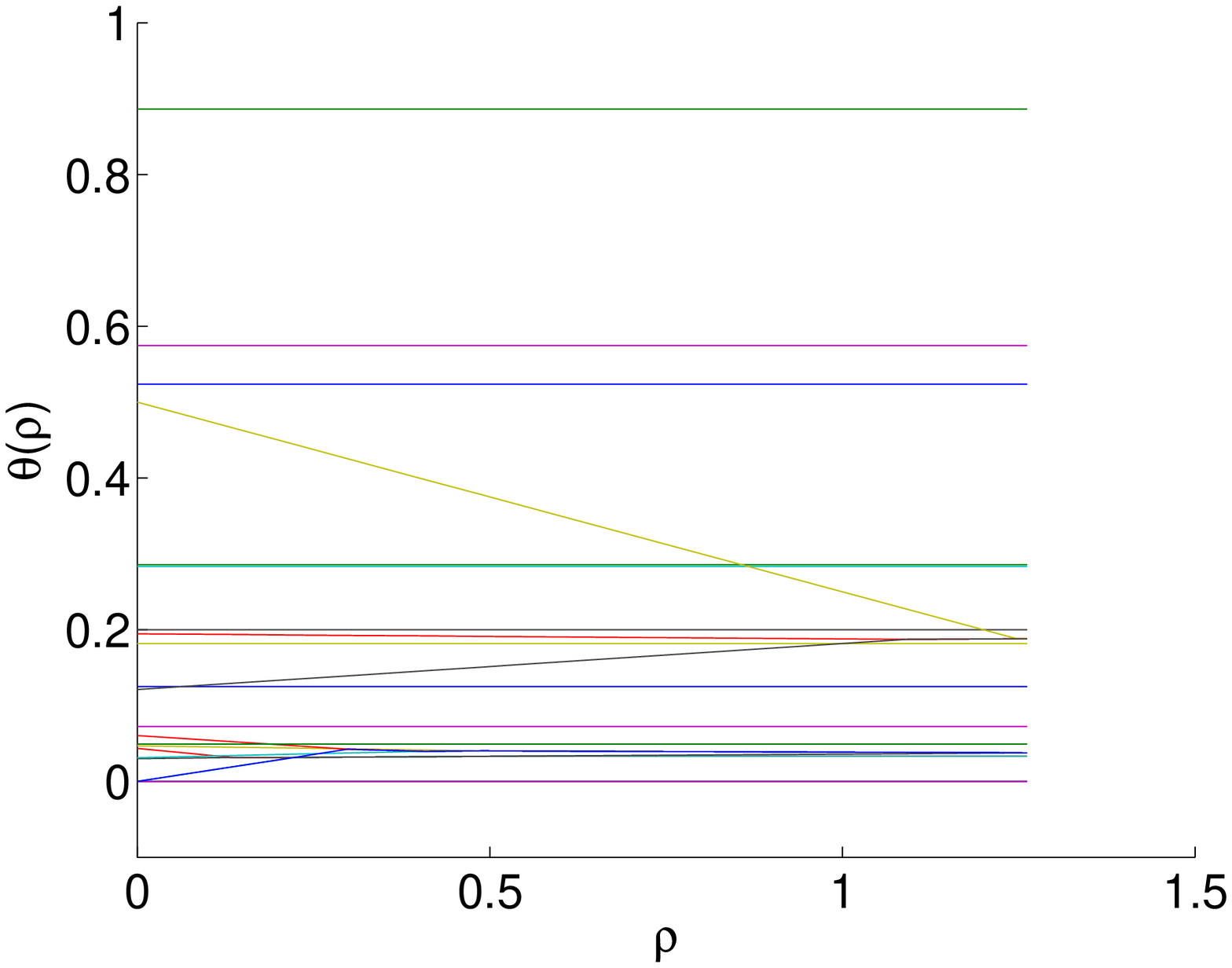} & \includegraphics[width=2.25in]{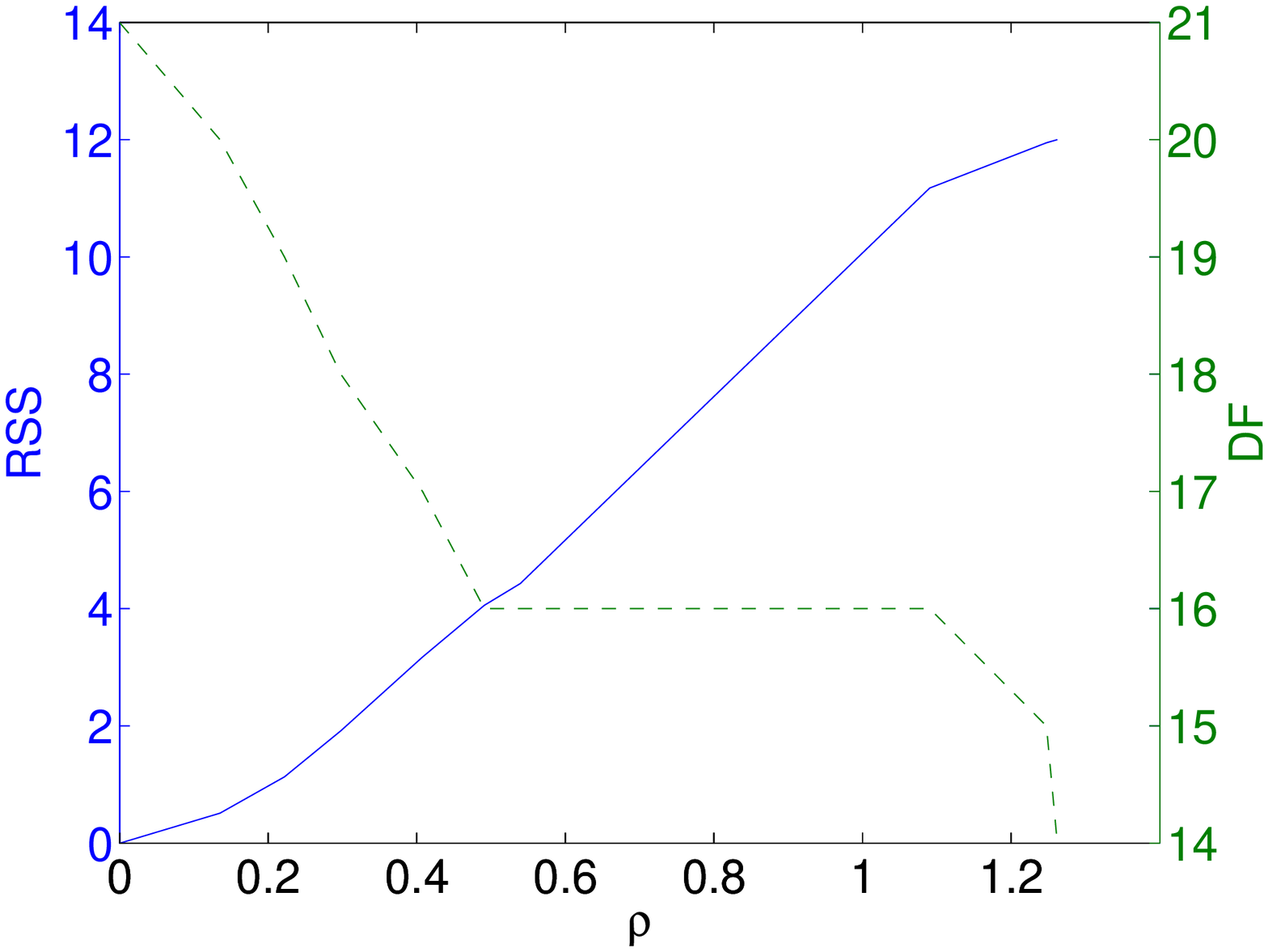}
\end{array}
$$
\caption{GPA prediction example. Left: Solution path for the predicted probabilities. Right: Residual sum of squares and effective degrees of freedom along the path.}
\label{fig:gpapred}
\end{figure}

The same authors \citep{RobertsonWrightDykstra88Book} predict the probability of obtaining a B or better college GPA based on high school GPA and ACT score. In their data covering 1490 college students, $\bar y_{ij}$ is the proportion of students who obtain a B or better college GPA among the $n_{ij}$ students who are within the $i$th ACT category and the $j$th high school GPA category. Prediction is achieved by minimizing the criterion $\sum_i \sum_j n_{ij} (\bar y_{ij} - \theta_{ij})^2$ subject to the matrix partial-order constraints $\theta_{11} \ge 0$, $\theta_{ij} \le \theta_{i+1,j}$, and $\theta_{ij} \le \theta_{i,j+1}$. Figure \ref{fig:gpapred} shows the solution path and the residual sum of squares and effective degrees of freedom along the path. The latter vividly illustrates the tradeoff between goodness of fit and degrees of freedom. Readers can consult page 33 of the reference
\citep{RobertsonWrightDykstra88Book} for the original data and the constrained parameter estimates.

\subsection{Nonparametric Shape-Restricted Regression }
\label{sec:nonpar}

In this section we visit a few problems amenable to the path algorithm arising in nonparametric statistics.  Given data $(x_i,y_i)$, $i=1,\ldots,n$, and a weight function $w(x)$, nonparametric least squares seeks a regression function $\theta(x)$ minimizing the criterion
\begin{eqnarray}
    \sum_{i=1}^n w(x_i) [y_i - \theta(x_i)]^2   \label{eqn:conc-reg-orig}
\end{eqnarray}
over a space ${\cal C}$ of functions with shape restrictions. In concave regression for instance, ${\cal C}$ is the space of concave functions. This seemingly intractable infinite dimensional problem can be simplified by minimizing the least squares criterion (\ref{eqn:concave-reg}) subject to inequality constraints. For a univariate predictor and concave regression, the constraints (\ref{eqn:concave-constraints}) are pertinent. The piecewise linear function extrapolated from the estimated $\theta_i$ is clearly concave. The consistency of concavity constrained least squares is proved by \cite{HansonPledger76ConcaveReg}; the asymptotic distribution of the corresponding estimator and its rate of convergence are investigated in later papers \citep{GroeneboomJongbloedWellner01ConvexReg,Mammen91Nonparam}. Other relevant shape restrictions for univariate predictors include monotonicity \citep{Brunk55Mono,Grenander56Mortality}, convexity \citep{GroeneboomJongbloedWellner01ConvexReg}, supermodularity \citep{Beresteanu04Nonpar}, and combinations of these.

Multidimensional nonparametric estimation is much harder because there is no natural order on $\mathbb{R}^d$ when $d > 1$. One fruitful approach to shape-restricted regression relies on sieve estimators \citep{Beresteanu04Nonpar,ShenWong94Sieve}. The general idea is to introduce a basis of local functions (for example, normalized B-splines) centered on the points of a grid $\bG$ spanning the support of the covariate vectors $\bx_i$. Admissible estimators are then limited to linear combinations of the basis functions subject to restrictions on the estimates at the grid points. Estimation can be formalized as minimization of the criterion $\|\by - \bPsi(\bX) \btheta\|_2^2$ subject to the constraints $\bC \bPsi(\bG) \btheta \le {\bf 0}$, where $\bPsi(\bX)$ is the matrix of basis functions evaluated at the covariate vectors $\bx_i$, $\bPsi(\bG)$ is the matrix of basis functions evaluated at the grid points, and $\btheta$ is a vector of regression coefficients.  The linear inequality constraints incorporated in the matrix $\bC$ reflect the required shape restrictions required. Estimation is performed on a sequence of grids (a sieve). Controlling the rate at which the sieve sequence converges yields a consistent estimator \citep{Beresteanu04Nonpar,ShenWong94Sieve}. Prediction reduces to interpolation, and the path algorithm
provides a computational engine for sieve estimation.

A related but different approach for multivariate convex regression minimizes the least squares criterion (\ref{eqn:concave-reg}) subject to the constraints $\bxi_i^t(\bx_j - \bx_i) \le \theta_j - \theta_i$ for every ordered pair $(i,j)$. In effect, $\theta_i$ is viewed as the value of the regression function $\theta(\bx)$ at the point $\bx_i$. The unknown vector $\bxi_i$ serves as a subgradient of $\theta(\bx)$ at $\bx_i$.  Because convexity is preserved by maxima, the formula
\begin{eqnarray*}
\theta(\bx) & = & \max_j \Big[\theta_j+\bxi_j^t (\bx - \bx_j) \Big]
\end{eqnarray*}
defines a convex function with value $\theta_i$ at $\bx=\bx_i$. In concave regression the opposite constraint inequalities are imposed.  Interpolation of predicted values in this model is accomplished by simply taking minima or maxima. Estimation reduces to a positive semidefinite quadratic program involving $n(d+1)$ variables and $n(n-1)$ inequality constraints. Note that the feasible region is nontrivial because setting all $\theta_i=0$ and all $\bxi_i={\bf 0}$ works. In implementing the extension of the path algorithm mentioned in Section \ref{path_extensions_section}, the large number of constraints may prove to be a hindrance and lead to very short path segments.  To improve estimation of the subgradients, it might be worth adding a small multiple of the ridge penalty $\sum_i \|\bxi_i\|_2^2$ to the objective function (\ref{eqn:concave-reg}). This would have the beneficial effect of turning a semidefinite quadratic program into a positive definite quadratic program.

\section{Conclusions}
\label{sec:conclusions}

Our new path algorithm for convex quadratic programming under affine constraints generalizes previous path algorithms for lasso penalized
regression and generalized lasso penalized regression.  By directly attacking the primal problem, the new algorithm avoids the circuitous tactic of solving the dual problem and translating the solution back to the primal problem. Our various examples confirm the path algorithm's versatility.  It's potential disadvantages involve computing the initial point $-\bA^{-1}\bb$ and storing the tableau.  In problems with large numbers of parameters, neither of these steps is trivial.  However, if $\bA$ has enough structure, then an explicit inverse may exist.  As we noted, once $\bA^{-1}$ is computed, there is no need to store the entire tableau. The multi-task regression problem with a large number of responses per case is a typical example where computation of $\bA$ simplifies.  In settings where the  matrix $\bA$ is singular, parameter constraints may compensate.  We have briefly indicated how to conduct path following in this
circumstance.

Our path algorithm qualifies as a general convex quadratic program solver. Custom algorithms have been developed for many special cases of quadratic programming. For example, the pool-adjacent-violators algorithm (PAVA) is now the standard approach to isotone regression
\citep{deLeeuwHornikMair09Isotone}. The other generic methods of quadratic programming include active set and interior point methods.  A comparison with our path algorithm would be illuminating, but in the interests of brevity we refrain from tackling the issue here. The path algorithm bears a stronger resemblance to the active set method. Indeed, both operate by deleting and adding constraints to a working active set.  However, the active set method must start with a feasible point, and interior point methods must start with points in the relative interior of the feasible region. The path algorithm's ability to deliver the whole regularized path with little additional computation cost beyond constrained estimation is bound to be appealing to statisticians.

\section*{Acknowledgements}
Kenneth Lange acknowledges support from United States Public Health Service grants GM53275, MH59490, CA87949, and CA16042.

\bibliography{../../../bib-HZ}
\bibliographystyle{Chicago}

\end{document}